\documentclass{aa}
\usepackage{graphics,psfig}
 
\newcommand{\msun}{\mbox{$M_\odot$}}

\newcommand{\usec}{\mbox{$\mu$s}}

\begin{document}      

\title{The parallax, mass and age of the PSR~J2145$-$0750
binary system}

\author{
        O.~L\"ohmer \inst{1},
        M.~Kramer \inst{2},
        T.~Driebe \inst{1},
        A.~Jessner \inst{1},
        D.~Mitra \inst{1},
        A.G.~Lyne \inst{2}
        }

\institute{Max-Planck-Institut f\"ur Radioastronomie, Auf dem H\"ugel 69,
              D-53121 Bonn, Germany
    \and
           University of Manchester, Jodrell Bank Observatory,
           Macclesfield, Cheshire SK11 9DL, UK
           }

\offprints{O.~L\"ohmer, e-mail: loehmer@mpifr-bonn.mpg.de}
\date{Received / Accepted}

\titlerunning{Parallax, mass and age of PSR~J2145$-$0750}
\authorrunning{O.~L\"ohmer et al.}

%%%%%%%%%%%%%%%%%%%%%%%%%%%%%%%%%%%%%%%%%%%%%%%%%%%%%%%%%%%%%%%%%%%%%
\abstract{ We present results of high-precision timing measurements of
  the binary millisecond pulsar \object{PSR J2145$-$0750}. Combining
  10~yrs of radio timing data obtained with the Effelsberg 100-m radio
  telescope and the Lovell 76-m radio telescope we measure a
  significant timing parallax of 2.0(6)\,mas placing the system at
  500~pc distance to the solar system. The detected secular change of
  the projected semi-major axis of the orbit $\dot x=1.8(6)\times
  10^{-14}$~lt-s~s$^{-1}$, where $x=(a_{\rm p}\sin i)/c$, is caused by
  the proper motion of the system. With this measurement we can
  constrain the orbital inclination angle to $i<61\degr$, with a
  median likelihood value of $46\degr$ which is consistent with
  results from polarimetric studies of the pulsar magnetosphere. This
  constraint together with the non-detection of Shapiro delay rules
  out certain combinations of the companion mass, $m_2$, and the
  inclination, $i$. For typical neutron star masses and using optical
  observations of the carbon/oxygen-core white dwarf we derive a mass
  range for the companion of $0.7\, M_\odot\leq m_2\leq 1.0\,
  M_\odot$. We apply evolutionary white dwarf cooling models to
  revisit the cooling age of the companion. Our analysis reveals that
  the companion has an effective temperature of $T_{\rm
  eff}=5750\pm600$~K and a cooling age of $\tau_{\rm
  cool}=3.6(2)$~Gyr, which is roughly a factor of three lower than the
  pulsar's characteristic age of 10.4 Gyr. The cooling age implies an
  initial spin period of $P_0=13.0(5)$~ms, which is very close to the
  current period.
\keywords{\it astrometry -- stars: neutron -- stars: white dwarfs
         -- binaries: general -- pulsars: individual (PSR
         J2145$-$0750) }}

\maketitle

%%%%%%%%%%%%%%%%%%%%%%%%%%%%%%%%%%%%%%%%%%%%%%%%%%%%%%%%%%%%%%%%%%%%%
\section{Introduction
\label{intro}}

Millisecond pulsars (MSPs) have small spin periods ($\sim$1$.5-30$~ms),
small spin-down rates ($\dot{P}\la 10^{-19}$ s s$^{-1}$), and
are believed to be recycled by the accretion of mass from an evolving
companion star (e.g.\ Alpar et al.\ 1982\nocite{acrs82}). About
$\sim$80\% of the MSP population in the Galactic plane are members of
binary systems. Most of those move in nearly circular orbits around
low-mass companions, forming the class of low-mass binary pulsars
(LMBPs). LMBPs have spin periods of $P\la 10$\,ms, small
eccentricities $e\la 10^{-3}$, and companions of mass
$0.15\,M_\odot\la m_2\la 0.4\,M_\odot$, presumably helium
(He)-core white dwarfs (WDs). The evolutionary history of LMBPs is
well understood. During the giant phase of the $\sim$1\,\msun\
companion star, mass overflows its Roche lobe and is accreted onto the
neutron star (NS). The stable mass accretion causes the pulsar to spin
up to millisecond periods and prevents a central helium ignition of
the giant, resulting in a binary system with a recycled millisecond
pulsar and a low-mass He-core WD (see review by Bhattacharya \& van
den Heuvel 1991\nocite{bv91}; Phinney \& Kulkarni 1994\nocite{pk94}).

In the last decade, pulsar surveys have revealed a new class of binary
pulsars, the intermediate-mass binary pulsars (IMBPs), with companion
masses above $0.45$\,\msun. Like in the LMBP case, the IMBPs have
almost circular orbits and WD companions, but show larger spin periods
$P\ga 10$~ms and slightly higher orbital eccentricities. As helium
ignition in the giant core starts above a core mass of 0.45\,\msun\
(Kippenhahn \& Weigert 1990\nocite{kw90}), the companions have to be
WDs with carbon/oxygen (CO) or oxygen/neon/magnesium (ONeMg)
cores. PSR J2145$-$0750 is most probably a member of the class of
IMBPs.  Discovered by Bailes et al.\ (1994\nocite{bhl+94}), the 16-ms
pulsar is in a nearly circular orbit, with an orbital period of 6.8
days.  The system shows some of the peculiar properties of the IMBP
class: (1) The pulsar companion has a mass of $m_2>0.43$\,\msun\ and
probably even larger than $0.51$\,\msun\ (50\% C.L.), suggesting it to
be a CO-core WD.  From optical observations of the PSR J2145$-$0750
system and using the dispersion measure (DM) distance of 500 pc,
Lundgren et al.\ (1996\nocite{lfc96}) concluded that the companion is
a WD. This is confirmed by our measurement of a timing parallax (see
Sect.~\ref{parallax}) showing that the DM distance is the true
distance of the pulsar.  (2) PSR J2145$-$0750 has a spin period of
16~ms, which is considerably longer than the average period of
$\sim$4~ms of LMBPs. (3) The surface magnetic field of the pulsar is
quite high: $B_{\rm S} =6\times 10^8$~G.  (4) Finally, the spin-down
age of the pulsar of $\sim$10~Gyr is very high and comparable to the
age of the Galaxy, raising interesting questions about its progenitor
and the initial spin period.

In contrast to LMBPs the evolution of IMBP progenitor systems is not
fully understood. Two models for the formation of heavy WDs in close
orbits have been proposed: (1) The system undergoes a Common Envelope
(CE) evolution and spiral-in phase on the Red Giant Branch (RGB)
(Tauris 1996\nocite{tau96}) or Asymptotic Giant Branch (AGB) (van den
Heuvel 1994\nocite{vdh94}), similar to the evolution of high-mass
binaries. (2) A massive companion progenitor looses much of its
envelope on the late main-sequence or early RGB through highly
super-Eddington mass transfer (Tauris et al.\ 2000\nocite{tvs00}; Taam
et al.\ 2000\nocite{tkr00}).

For the particular case of PSR J2145$-$0750, van den Heuvel (1994)
proposed an evolutionary channel with a CE spiral-in phase on the
AGB. In this scenario the progenitor system consisted of a
$1-6$\,\msun\ donor star, with an orbital period such that this star
overflows its Roche lobe on the AGB.  During the CE phase, which is
similar to that of high-mass binary pulsar progenitors, the mass
accretion onto the pulsar is highly unstable. This results in an
incomplete recycling process of the pulsar and can explain the high
values for $P$ and $B_{\rm S}$. One possibility to test these
evolutionary channels is to measure the NS and companion masses. Here,
we report on new results from timing measurements of the PSR
J2145$-$0750 binary system. They allow us to estimate the inclination
and companion mass of the binary. From the detection of a timing
parallax we derive an independent distance estimate of the pulsar.

For a pulsar - white dwarf binary system one usually expects that the
cooling age of the WD matches the age of the pulsar. This is explained
by the fact that the cooling and spin-down clocks start ticking at
approximately the same time when the companion starts to contract to a
WD and the pulsar turns on at the end of the mass transfer phase. The
cooling age of the WD companion can be determined from its effective
temperature and its mass, using a WD cooling model.  We use the
parallax distance along with optical observations to revisit the
cooling age of the WD companion, based on evolutionary WD cooling
models. We discuss the implications of our findings for the
evolutionary history of the binary.

%%%%%%%%%%%%%%%%%%%%%%%%%%%%%%%%%%%%%%%%%%%%%%%%%%%%%%%%%%%%%%%%%%%%%
\section{Observations\label{obs}}

We have made regular high-precision timing observations of
PSR~J2145$-$0750 over a 10~yrs time span using both the 100-m
Effelsberg radio-telescope of the Max-Planck-Institut f\"ur
Radioastronomie in Bonn, Germany, and the 76-m Lovell radio-telescope
at Jodrell Bank, UK.

The Effelsberg observations have been carried out since April 1994 at
a centre frequency of 1410~MHz. Data have been acquired approximately
once per month, with a typical observing time of $3\times 7$~min. For
the observations we used a 1300$-$1700~MHz tunable HEMT receiver with
a system temperature of $30$~K on the cold sky and an antenna gain of
1.5~K~Jy$^{-1}$. In order to monitor changes of the dispersion measure
(DM) we occasionally collected data at 860~MHz, using an uncooled HEMT
receiver with a system temperature of $65$~K on the cold sky and a
gain of 1.5~K~Jy$^{-1}$.  During the period April 1994 -- October
1996, the data were obtained using the Effelsberg Pulsar Observation
System (EPOS) (Seiradakis et al.\ 1995\nocite{sgg+95}; Kramer et
al.\ 1997, 1998\nocite{kxj+97}\nocite{kxl+98}). Here, the two
circular polarization signals were processed in a $4\times 60 \times
666$~kHz filter bank combined with an incoherent hardware
de-disperser. The detected left- and right-hand circular signals in
each channel were de-dispersed on-line, added, and folded with the
topocentric pulse period to produce a total power pulse profile of
40~MHz bandwidth.  Since October 1996 the observations were made with
the Effelsberg--Berkeley Pulsar Processor (EBPP), which corrects for
the dispersion smearing of the signal using a coherent de-dispersion
technique (Hankins \& Rickett 1975\nocite{hr75}). In the total power
mode the EBPP provides 32 channels for both polarizations with a
maximum total bandwidth of 112~MHz, depending on DM and observing
frequency (Backer et al.\ 1997\nocite{bdz+97}). For PSR~J2145$-$0750 a
total bandwidth of 90~MHz was available at 1410~MHz. The output
signals of each channel were fed into de-disperser boards for coherent
on-line de-dispersion and were synchronously folded at the pulse
period over 7-min integrations. To obtain a high signal-to-noise ratio
polarization profile of PSR J2145$-$0750 (see Sect.~\ref{pol}) we used
the EBPP in polarization mode where the four Stokes parameters are
available over 32 channels with a total bandwidth of 28~MHz.

Lovell observations of PSR J2145$-$0750 were made from October 1992 to
January 1999 at centre frequencies of 606 and 1400 MHz. Both circular
polarization signals were detected and incoherently de-dispersed in a
$2\times 64\times 0.125$-MHz filter bank at 606 MHz and in a $4\times
32\times 1.0$-MHz filter bank at 1400 MHz. The signals were added and
synchronously folded at the pulse period with a typical integration
time of $1-3$~min.  Details of the observing system can be found in
Gould \& Lyne (1998)\nocite{gl98}.

Both Effelsberg and Lovell data were time stamped with clock
information from a local hydrogen maser clock and later synchronized
to UTC(NIST) using the signals from the Global Positioning System
(GPS). In order to calculate the pulse time-of-arrival (TOA) synthetic
templates of the pulse profile were constructed for each backend and
observing frequency, and fitted to the observed profiles with template
matching procedures (details can be found in Lange et al.\
2001\nocite{lcw+01}). TOA uncertainties were estimated using a method
described by Downs \& Reichley (1983)\nocite{dr83}. Typical TOA errors
for EBPP observations are 2.8\,$\mu$s at 860~MHz and
1.9\,$\mu$s at 1410~MHz, whereas for Lovell observations we found
5.2\,$\mu$s at 606~MHz and 4.5\,$\mu$s at 1400~MHz.

%%%%%%%%%%%%%%%%%%%%%%%%%%%%%%%%%%%%%%%%%%%%%%%%%%%%%%%%%%%%%%%%%%%%%
\section{Timing analysis \label{timanalysis}}

The combined TOAs, weighted by their individual uncertainties, were
fitted to a spin-down model of the pulsar in a binary system with the
software package {\sc
tempo}\footnote{http://pulsar.princeton.edu/tempo}, using the DE200
planetary ephemerides (Standish 1990\nocite{sta90}). PSR J2145$-$0750
lies close to the ecliptic plane, with an ecliptic latitude of only
$\beta=5^{\circ}\!.3$, resulting in less accurate position and proper
motion measurements if determined in the equatorial reference
frame. In order to minimize covariances between the astrometric
parameters we chose ecliptic coordinates for our timing model,
performing standard Monte-Carlo simulations to derive reliable values
for the proper motion $\mu_\lambda$ and $\mu_\beta$.  Since the pulsar
is moving in an almost circular orbit we applied the timing model for
binary systems with small eccentricities (ELL1) using the
Laplace-Lagrange parameters $\epsilon_1$ and $\epsilon_2$, as well as
the time of ascending node $T_{\rm ASC}$ (Lange et al. 2001). The
three Keplerian parameters eccentricity $e$, epoch $T_{\rm 0}$, and
longitude of periastron $\omega$, are then calculated from the former
three parameters.

In the fitting procedure the TOA segments obtained with EPOS, EBPP,
and Lovell data were fitted for a mutual offset accounting for
different templates and TOA reference points in the profiles. Using
the full TOA data set at all frequencies we then determined the DM
over the 10-yrs period. By excluding all TOAs with lines-of sight to
the pulsar lying closer than 30$\degr$\ to the Sun we analysed the
influence of free electrons in the solar system on the DM, but found
no variations of the DM or other parameters larger than their
1$\sigma$ uncertainties.  However, the multi-frequency timing fit
resulted in a detection of a secular DM variation of $d({\rm DM})/dt
=-2.2(4)\times 10^{-4}$~pc cm$^{-3}$ yr$^{-1}$. Variations of pulsar
DMs have been explained by a simple wedge model of electron density
fluctuations in the Galaxy, resulting in a square root dependence of
$d({\rm DM})/dt\propto \sqrt{\rm DM}$ (Backer et al.\
1993\nocite{bhh93}).  Recently, Hobbs et al.\ (2004) showed that this
relation holds for a large group of pulsars, and derived a best-fit of
$d({\rm DM})/dt\approx 0.0002\,\sqrt{\rm DM}$, with a scatter of one
order of magnitude. For PSR J2145$-$0750 our measurement of $|d({\rm
DM})/dt|$ is consistent with the empirical relation. By holding the
values for DM and $d({\rm DM})/dt$ fixed we then obtained the best-fit
timing model for the astrometric, rotational and binary parameters
using only the more accurate 1400~MHz TOAs from Effelsberg and Jodrell
Bank. We scaled the TOA errors by an appropriate factor to achieve a
uniform $\chi^2/n_{\rm dof}=1$ for each TOA segment.
%
%%%%%%%%%%%%%%%%%%%%%%%%%%%% FIGURE %%%%%%%%%%%%%%%%%%%%%%%%%%%%%%%%%%
\begin{figure}
\centering
\psfig{file=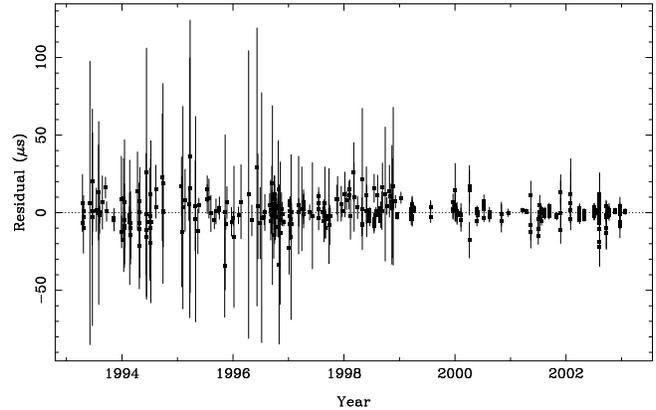,angle=-90,width=8.5cm}
\caption{ \label{fig:resid}
Post-fit timing residuals of the combined Effelsberg and Lovell 1400~MHz 
timing data as a function of observing year.
}
\end{figure}
%
%%%%%%%%%%%%%%%%%%%%%%%%%%%% FIGURE %%%%%%%%%%%%%%%%%%%%%%%%%%%%%%%%%%
\begin{figure}
\centering
\psfig{file=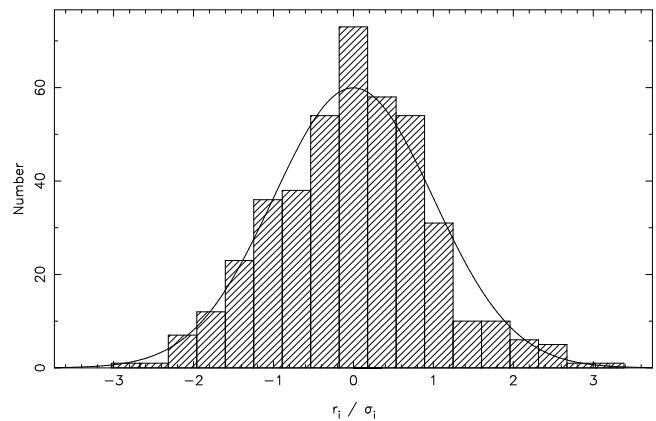,angle=-90,width=8.5cm}
\caption{ \label{fig:hist}
Distribution of the 1400~MHz post-fit timing residuals normalized by their
uncertainties.
}
\end{figure}

From the best-fit model we obtained post-fit timing residuals that are
displayed in Fig.~\ref{fig:resid}. As seen from the figure, we obtained
a good coverage of the 10-yrs timing baseline. Note that the EPOS and
Lovell TOAs obtained from 1992 -- 1999 have much higher uncertainties
than the EBPP data which is due to the incoherent dispersion removal
technique. In order to test the statistical properties of our data
we generated a histogram of the deviations of the timing residuals
from the model as shown in Fig.~\ref{fig:hist}.  The post-fit residuals
are consistent with the expected Gaussian distribution.  The timing
parameters of the best-fit model for PSR~J2145$-$0750 are listed in
Table~\ref{tab:params}. Included in the Table are upper limits as well
as derived parameters. The limits on the pulse frequency second
derivative, $\ddot\nu$, and the orbital period derivative, $\dot
P_{\rm b}$, were found by allowing the extra parameters to vary one at
a time in the global fit. Fitting for the derivatives of both the
Laplace-Lagrange parameters, $\dot\epsilon_1$ and $\dot\epsilon_2$,
led to the upper limits on the eccentricity derivative, $\dot e$, and
the periastron rate of change, $\dot\omega$.
%
%%%%%%%%%%%%%%%%%%%%%%%%%%%% Table 1 %%%%%%%%%%%%%%%%%%%%%%%%%%%%%%%%%%
\begin{table*}
\begin{center}
\caption{Timing model for PSR~J2145$-$0750\label{tab:params}
}
\vspace{0.1cm}
\begin{tabular}{ll}
\hline\hline
\multicolumn{2}{c}{\it Measured parameters$^a$}\\
\hline
Ecliptic longitude, $\lambda$       & $326\fdg02465688(10)$\\
Ecliptic latitude, $\beta$         & $5\fdg3130779(8)$\\
Proper motion$^b$, $\mu_{\lambda}$ (mas\,yr$^{-1}$)   & $-12.37(6)$ \\
Proper motion$^b$, $\mu_{\beta}$   (mas\,yr$^{-1}$)   & $-6.8(7)$ \\
Right ascension$^c$, $\alpha$ (J2000) & $21^\mathrm{h} 45^\mathrm{m} 50\,\fs46726$ \\
Declination$^c$, $\delta$ (J2000)     & $-07\degr 50\arcmin 18\,\farcs 375$\\
Parallax, $\pi$ (mas)                     & $2.0(6)$ \\
Pulse frequency, $\nu$ (s$^{-1}$)          & $62.2958888453343(4)$\\
Pulse frequency derivative, $\dot\nu$ (10$^{-16}$ s$^{-2}$) & $-1.15481(8)$\\
Pulse period, $P$ (ms)                     & $16.05242365965367(14)$\\
Period derivative, $\dot P$ (10$^{-20}$ s\,s$^{-1}$) & $2.9757(2)$\\
Epoch (MJD)                              & $50800.0$ \\
Orbital period, $P_{\rm b}$ (d)          & $6.8389025099(4)$\\
Projected semi-major axis, $x$ (lt-s)    & $10.1641056(6)$    \\
First derivative of $x$, $\dot x$ (10$^{-14}$ lt-s~s$^{-1}$)& $1.8(6)$\\
$\epsilon_1$          & $-0.00000677(8)$      \\
$\epsilon_2$          & $-0.00001805(10)$      \\
Epoch of ascending node, $T_{\rm ASC}$ (MJD)   & $50802.29811051(8)$\\
Eccentricity$^d$, $e$                    & $0.00001928(10)$      \\
Epoch of periastron$^d$, $T_{\rm 0}$ (MJD)      & $50806.108(4)$\\
Longitude of periastron$^c$, $\omega$ (deg)     & $200.5(2)$    \\
Dispersion measure, DM (pc\,cm$^{-3}$) & $9.0031(2)$  \\
Dispersion measure derivative, $d({\rm DM})/dt$ (10$^{-4}$ pc\,cm$^{-3}$\,yr$^{-1}$) & $-2.2(4)$  \\
\hline
\multicolumn{2}{c}{\it Measured upper limits$^e$}\\
\hline
Pulse frequency second derivative, $|\ddot\nu|$ (s$^{-3}$)& $<8\times 10^{-28}$\\
Orbital period derivative, $|\dot P_{\rm b}|$ (s\,s$^{-1}$)& $<2.4\times 10^{-12}$ \\
Eccentricity derivative, $|\dot e|$ (s$^{-1}$)     & $<2.8\times 10^{-15}$   \\
Periastron rate of change, $|\dot\omega|$ (deg\,yr$^{-1}$) & $<0.47$ \\ 
\hline
\multicolumn{2}{c}{\it Derived parameters}\\
\hline
Galactic longitude, $l$  & $47\fdg 78$\\
Galactic latitude, $b$ & $-42\fdg 08$\\
Parallax distance  (pc)& $500^{+210}_{-120}$  \\
Composite proper motion, $\mu$ (mas\,yr$^{-1}$)   & 14.1(4)\\
Transverse velocity, $v_{\rm t}$ (km\,s$^{-1}$)  &  33(9) \\
Mass function, $f_{\rm m}$ (\msun)     & $0.024105570(4)$\\
Pulsar characteristic age (Gyr) & 10.4(5)\\
Pulsar surface magnetic field ($10^8$ G) & 6.35(14)\\
Timing RMS (\usec)           & 2.7        \\
\noalign{\smallskip}\hline
\end{tabular}
\end{center}
{\small $^a$ Uncertainties quoted are in the last digit(s) and represent
  $2\sigma$ estimates (twice the formal {\sc tempo} errors)}\newline
{\small $^b$ calculated from Monte-Carlo simulations}\newline
{\small $^c$ calculated from ecliptic coordinates}\newline
{\small $^d$ calculated from the ELL1 model}\newline
{\small $^e$ Upper limits represent 95\% C.L.}\newline
\end{table*}
%%%%%%%%%%%%%%%%%%%%%%%%%%%%%%%%%%%%%%%%%%%%%%%%%%%%%%%%%%%%%%%%%%%%%

%%%%%%%%%%%%%%%%%%%%%%%%%%%%%%%%%%%%%%%%%%%%%%%%%%%%%%%%%%%%%%%%%%%%%
\section{Astrometry and intrinsic pulsar parameters  \label{astrom}}

\subsection{Parallax, proper motion and transverse velocity
\label{parallax}}

Pulsar distances are estimated from their DM using models for the
electron density distribution in the Galaxy (e.g. Taylor \& Cordes
1993\nocite{tc93}, updated model: Cordes \& Lazio 2002\nocite{cl02a},
hereafter NE2001). Independent distance measurements define an
absolute distance scale and hence are of fundamental importance to
calibrate models like NE2001. For pulsars, distances are derived from
trigonometric parallaxes using VLBI observations, or from pulsar
timing.  In pulsar timing, the parallax is obtained by measuring the
time delay of the TOA caused by the curvature of the radio wave
fronts.  However, the amplitude of the corresponding residual term,
$\Delta t_\pi\: (\mu{\rm s}) = a^2 \cos\beta\,/\,(2cd) =
1.21\cos\beta\,/\,d$, is very small, allowing us to detect a timing
parallax only for a very few millisecond pulsars near the ecliptic
plane.  Here, $a$ is the Earth-Sun distance in kpc, $\beta$ the
ecliptic latitude, and $d$ the distance to the pulsar in kpc. Timing
parallaxes have been detected for pulsars such as \object{PSR
B1855+09} (Kaspi et al.\ 1991\nocite{ktr94}), \object{PSR J1713+0747}
(Camilo et al.\ 1994a\nocite{cfw94}), \object{PSR J0437$-$4715}
(Sandhu et al.\ 1997\nocite{sbm+97}) and \object{PSR J1744$-$1134}
(Toscano et al.\ 1999\nocite{tsb+99}). Nevertheless, due to an
ecliptic latitude of only $\beta=5^{\circ}\! .3$ we were able to
detect for the first time a timing parallax for PSR~J2145$-$0750 of
$\pi=2.0(6)$~mas. This corresponds to a distance estimate of
$d_\pi=500^{+210}_{-120}$~pc and agrees very well with the DM
distances, i.e.\ $d_{\rm DM}=500$~pc from the Taylor \& Cordes (1993)
model and $d_{\rm DM}=570$~pc from the NE2001 model (with typical
uncertainties of $\sim$10\%). Using the parallax distance we derive a
transverse velocity of $v_{\rm t}=33(9)$~km\,s$^{-1}$, which is in
excellent agreement with scintillation velocities of $31\pm 25$~km
s$^{-1}$ obtained by Nicastro \& Johnston (1995\nocite{nj95}) and
Johnston et al.\ (1998\nocite{jnk98}).

With a galactic position angle of 289$\degr$ the pulsar is moving
towards the Galactic plane.  This is not in contradiction to the idea
that pulsars are born in the Galactic plane and then move to higher
latitudes, as the MSP population is very old ($\sim$1~Gyr) and has
already reached a dynamic equilibrium in the Galactic gravitational
potential (e.g.\ Toscano et al.\ 1999). In this scenario PSR
J2145$-$0750, with a characteristic age of 10~Gyr (see Sect.
\ref{age}), would have already performed many oscillations in the
Galactic potential described by Kuijken \& Gilmore (1989).

%%%%%%%%%%%%%%%%%%%%%%%%%%%%%%%%%%%%%%%%%%%%%%%%%%%%%%%%%%%%%%%%%%%%%
\subsection{Doppler corrections}
\label{doppler}

Accelerations in the Galactic gravitational potential introduce
contributions to the observed period derivative $\dot P$ which cannot
be neglected for MSPs that have spin-down rates of about six orders of
magnitude lower than those of normal pulsars. Doppler effects arise
due to (a) the Galactic differential rotation, (b) the vertical
acceleration in the Galaxy (Damour \& Taylor 1991\nocite{dt91}), and
(c) the transverse velocity of the pulsar (Shklovskii
1970\nocite{shk70}). Following these authors, the Doppler correction
to the period derivative can be written as
\begin{equation}\label{dopplercorr}
\left(\frac{\dot P}{P}\right)^{\rm D} = 
\frac{a_z\sin b}{c}
\,-\,\frac{v_0^2}{c\,R_0}\cos b\left(\cos\, l + \frac{\beta}
{\sin^2 l + \beta^2}\right)  \,+\, \frac{\mu^2\,d}{c}\ ,
\end{equation}
where $R_0$ and $v_0$ denote the galactocentric radius and galactic
circular velocity of the Sun, respectively, $l$ and $b$ the galactic
longitude and latitude of the pulsar, and $\beta=d/R_0-\cos\, l$. We
choose $R_0=8.0$~kpc and $v_0= 220$~km~s$^{-1}$ (Reid
1993\nocite{rei93}).  $a_z$ is the vertical component of the galactic
acceleration and can be calculated from the Kuijken \& Gilmore
(1989\nocite{kg89}) model of the Galactic potential. For
PSR~J2145$-$0750 we derive a kinematic correction of $(\dot P/P)^{\rm
D} = 3.3(7)\times 10^{-19}$~s$^{-1}$, where the major contribution
(73\%) comes from the Shlovskii term $(\mu^2\,d)/c$.

%%%%%%%%%%%%%%%%%%%%%%%%%%%%%%%%%%%%%%%%%%%%%%%%%%%%%%%%%%%%%%%%%%%%%
\subsection{Characteristic age and magnetic field
\label{age}}

Subtracting the kinematic contributions from the observed spin-down rate
of PSR~J2145$-$0750 we obtain an intrinsic spin-down rate of $\dot
P^{\rm intr}=\dot P - \dot P^{\rm D}=2.45(11)\times 10^{-20}$~s
s$^{-1}$ which is 18\% smaller than the observed value.

It is usually assumed that the evolution of the spin frequency
$\nu=1/P$ can be described by a power-law $\dot\nu\propto -\nu^n$,
where $n$ is the so-called {\it braking index}. Then the pulsar age
can be calculated from
\begin{equation}\label{pulsarage}
\tau = \frac{P}{(n-1)\,\dot P}\:
\left[1-\,\left(\frac{P_0}{P}\right)^{n-1}\right]\ \ \ ,
\end{equation}
where $P_0$ is the initial period of the pulsar. Under the assumptions
that $P_0\ll P$ and that the spin-down is due to magnetic dipole
braking ($n=3$), the spin-down age is given by $\tau_{\rm c} =
P/(2\,\dot P)$, the {\em characteristic age} of the pulsar.
Subtracting the Doppler contributions, for PSR~J2145$-$0750 we thus
obtain $\tau_{\rm c} = P/(2\,\dot P^{\rm intr}) = 10.4(5)$ Gyr.  As
the uncertainty of $\tau_{\rm c}$ caused by the a priori unknown $n$
and $P_0$ can be considerable, an independent age estimate, as given
by the cooling age of the pulsar companion (see Sect.~\ref{wdcool}),
is highly desirable.

Assuming a pure dipolar magnetic field and using the intrinsic
spin-down rate we can estimate the surface magnetic field of the
pulsar to be $B_{\rm S} = 3.2\times 10^{19}\sqrt{P\dot P^{\rm
intr}}$~G = 6.35(14)$\times 10^8$~G. This magnetic field is more than
three orders of magnitude smaller than for normal pulsars, a clear
indication than PSR J2145$-$0750 has gone through a recycling phase.

%%%%%%%%%%%%%%%%%%%%%%%%%%%%%%%%%%%%%%%%%%%%%%%%%%%%%%%%%%%%%%%%%%%%%
\subsection{Timing instabilities}

Any remaining non-gaussian noise in timing residuals is generally
explained by rotational instabilities intrinsic to the pulsar, or by
an incomplete timing model, such as unmodelled DM variations or the
existence of a planet in the system. For PSR J2145$-$0750 a fit for
$\ddot\nu$ over the full timing baseline resulted in
$\ddot\nu=-5(3)\times 10^{-28}$~s$^{-3}$. As this measurement is
hardly significant we conservatively quote $\ddot\nu$ as an upper
limit in Table~\ref{tab:params}.  At the same time, we do not detect
any DM variations that are not accounted for in the timing model. A
simultaneous fit of our multi-frequency data for third- and
fourth-order derivatives of DM as well as for $\ddot\nu$ resulted in a
similar $\ddot\nu$ value, while $d^2({\rm DM})/dt^2$ and $d^3({\rm
DM})/dt^3$ were not significant.

Arzoumanian et al.\ (1994\nocite{antt94}) analysed the timing noise of
a number of slow and millisecond pulsars by quantifying the noisiness
with the noise parameter $\Delta_8 = \log{(|\ddot\nu|\,t^3/(6\nu))}$,
where $t=10^8$~s. They quote a fit-by-eye to the data of the form
$\Delta_8 = 6.6+0.6\log\dot P$, with a large scatter of the data
points. Lommen (2002\nocite{lom02}) revised this relation using new
MSP data and showed that at least seven MSPs have significantly lower
$\Delta_8$'s than presented by Arzoumanian et al.\ (1994).

In order to estimate the timing noise of PSR J2145$-$0750 we divided
the total data set into three sub-intervals, each spanning $10^8$~s
$\approx 3.2$~yr of TOAs.  For each sub-interval we transformed
epoch, pulse frequency $\nu$, and epoch of ascending node $T_{\rm asc}$
to an epoch near the centre of each sub-interval. We then fitted the
timing model of the pulsar, holding DM and $d({\rm DM})/dt$ fixed and
allowing $\ddot\nu$ to vary. The resulting $\Delta_8$'s for each
sub-interval are $-4.0$, $-4.1$, and $-4.3$, i.e.\ slightly larger than
the predicted value of $\Delta_8=-5.1$. We thus conclude that PSR
J2145$-$0750 is a possible candidate for intrinsic timing
noise. Future timing observations will resolve this assertion.

%%%%%%%%%%%%%%%%%%%%%%%%%%%%%%%%%%%%%%%%%%%%%%%%%%%%%%%%%%%%%%%%%%%%%
\subsection{Secular changes in orbital parameters\label{secchange}}

For the first time we detected a secular variation of the projected
semi-major axis of PSR~J2145$-$0750 of $\dot x=1.8(6)\times
10^{-14}$~lt-s~s$^{-1}$. In order to test the robustness of this
measurement we divided the total set into five sub-intervals, each
spanning two years of TOAs. For every sub-interval the TOAs were
fitted to the timing model of the pulsar, with the variation of $x$
held fixed at zero. Hereby we transformed epoch, pulse frequency $\nu$,
and epoch of ascending node $T_{\rm asc}$ to an epoch near the centre
of each sub-interval. The fits resulted in an increase of $x$ with
time, with a slope of $1.7(9)\times 10^{-14}$~lt-s~s$^{-1}$, confirming
the value from the global fit.

The timing data showed no evidence for secular changes in other
orbital parameters. In principle, the binary system could undergo both a
general relativistic (GR) advance of periastron and an orbital decay
due to the emission of gravitational radiation (Damour \& Taylor
1991\nocite{dt91}), i.e.\
\begin{eqnarray}
\dot \omega^{\rm GR} &=& 3\left(\frac{P_{\rm b}}{2\pi}\right)^{-5/3}
             \frac{1}{1-e^2}\left(T_{\odot} M\right)^{2/3}
\label{omdot-gr}\\
\noalign{\medskip}
\dot P_{\rm b}^{\rm GR} &=& -\frac{192\pi}{5}
       \left(\frac{P_{\rm b}}{2\pi}\right)^{-5/3}
       \frac{( 1+\frac{73}{24}e^2+\frac{37}{96}e^4)}{(1-e^2)^{7/2}}
\nonumber\\
     & &   \times\ T_{\odot}^{5/3} m_1 m_2 M^{-1/3} 
\label{pbdot-gr} 
\end{eqnarray}
where the masses are in Solar units and $T_{\odot}=(G\,M_{\odot}/c^3)$.  

If we assume a pulsar mass of $m_1=1.4$\,\msun\ and a companion mass
of $m_2=0.51$\,\msun\ (using the mass function and a median
inclination $<i>=60\degr$), we expect $\dot\omega^{\rm GR} =
0^{\circ}\!  .0124$~yr$^{-1}$ and $\dot P_{\rm b}^{\rm GR}\sim
-5.1\times 10^{-16}$~s~s$^{-1}$. However, as PSR~J2145$-$0750 has a
very small eccentricity, all effects of secular variations in $\omega$
are fully absorbed by the redefinition of the binary period and are
therefore not observable in this system.  The observed upper limit for
$\dot P_{\rm b}$ is four orders of magnitude higher which makes any
detection of the GR orbital decay highly unlikely.

Moreover, the Shklovskii effect discussed in Sect.~\ref{doppler}
also affects the observed value of $\dot P_{\rm b}$.  Using the
results from Sect.~\ref{doppler}, we expect a contribution of
$\dot{P}_{\rm b}^{\rm Shk}= 2.4(5)\times 10^{-19}\ {\rm s}^{-1}\times
P_{\rm b} = 1.4(3)\times 10^{-13}$~s~s$^{-1}$. This value is not only
of different sign as the expected GR contribution but also about three
orders of magnitude larger. Nevertheless, the Shklovskii term is
still consistent with the observations as it is yet a factor of 20
smaller than our derived upper limited.

%%%%%%%%%%%%%%%%%%%%%%%%%%%%%%%%%%%%%%%%%%%%%%%%%%%%%%%%%%%%%%%%%%%%%
\section{Orbital inclination and mass}

\subsection{Secular change of x\label{xdot}}

The observed secular change of the projected semi-major axis $x=a_{\rm
p}\sin i/c$ could, in principle, result from a variation of the major
axis of the pulsar orbit, $a_{\rm p}$, of the orbital inclination,
$i$, or a combination of both.  The contributions to $\dot x$ are
(Damour \& Taylor 1992\nocite{dt92}; Kopeikin 1996\nocite{kop96}):
\begin{equation}
    \left(\frac{\dot x}{x} \right)^{\rm obs}\ =\
    \left(\frac{\dot a_{\rm p}}{a_{\rm p}}\right)^{\rm GW} +
    \left(\frac{\dot x}{x}\right)^{\rm D} +
    \left(\frac{\dot x}{x}\right)^{\rm PM}\ \ \ .
\end{equation}

The emission of gravitational waves (GW) leads to a shrinking of the
pulsar orbit given by $(\dot a_{\rm p}/a_{\rm p})^{\rm GW}=2/3\times
(\dot P_{\rm b}/P_{\rm b})^{\rm GW}$ (e.g.\ Doroshenko et al.\
2001\nocite{dlk+01}).  With Eq.~\ref{pbdot-gr} one expects $|\dot
a_{\rm p}/{a_{\rm p}}|^{\rm GW} \sim 5.8\times 10^{-22}$\,s$^{-1}$
which is seven orders of magnitude smaller than the observed $|\dot
x/x|^{\rm obs}=1.8\times 10^{-15}$\,s$^{-1}$. In general, $|\dot
a_{\rm p}/{a_{\rm p}}|$ will be of the same order of magnitude as
$|\dot P_{\rm b}/P_{\rm b}|$ for typical astrophysical processes in a
binary. With the observed upper limit of $\dot P_{\rm b}<2.4\times
10^{-12}$\,s$^{-1}$ we derive $|\dot P_{\rm b}/P_{\rm b}|<4.1\times
10^{-18}$\,s$^{-1}$, still three orders of magnitude smaller than
$|\dot x/x|^{\rm obs}$. Therefore, the nonzero $\dot x$ is most likely
not caused by orbital evolution.

The Doppler correction affects the light-travel time across the orbit
in the same way as the pulse period, so that the term $(\dot x/x)^{\rm
D}$ is identical to the one for the period derivative
(Eq.~\ref{dopplercorr}, see Damour \& Taylor 1992). Its value is
therefore four orders of magnitude smaller than the observed value
$(\dot x/x)^{\rm obs}$.

We conclude that the nonzero $\dot x$ must arise from a change
in the observed inclination $i$ of the orbit. The relative motion of
the binary and the observer causes a change of the orientation of the
line-of-sight to the binary which results in an apparent change of
$i$. Kopeikin (1996) derived
\begin{equation}\label{xdotpm}
\left(\frac{\dot x}{x}\right)^{\rm PM} = \mu\,\cot i \,\sin\theta \ \
\ ,
\end{equation}
where $\theta$ denotes the difference of the position angle of proper
motion and the position angle of ascending node. As $\theta$ is a
priori unknown, we can only derive a firm upper limit for $i$ using
$|\sin\theta|<1$ and the lower limit of allowed values for $\dot x$
from Table~\ref{tab:params}, i.e.\ $i < \tan^{-1}(\mu\,x/\dot x_{\rm
min}) = 61^\circ$.

\begin{figure}
\centering
\psfig{file=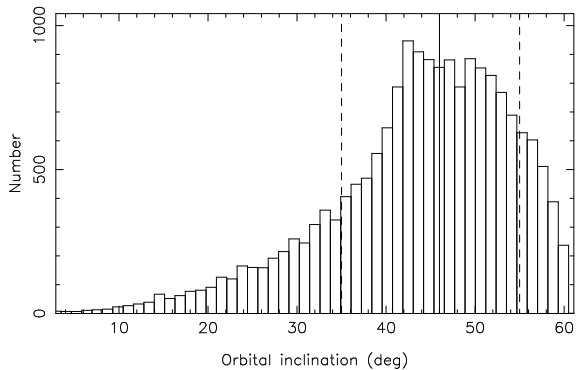,angle=-90,width=7.5cm}
\caption{\label{fig:mcfit-incl}
  Distribution of orbital inclinations $i$ resulting from Monte-Carlo
  simulations (see Sect.~\ref{xdot}). The median of the distribution is
  $46\degr$ (solid line), and its lower and upper 1$\sigma$ values are at
  $35\degr$ and $55\degr$, respectively (dashed lines).  }
\end{figure}

We performed Monte-Carlo simulations to derive the distribution of
inclination angles within this constraint, following a procedure
described by Nice et al.\ (2001\nocite{nss01}).  Assuming that the
orientation of the binary in space is arbitrary, $\theta$ is a
uniformly distributed random variable and $i$ is a random variable
distributed with uniform probability in $\cos i$.  From these
distributions we select values of $\theta$ and $i$ and retain only
those combinations which satisfy Eq.~\ref{xdotpm} within the
measurement errors. The resulting distribution for $i$ is presented in
Fig.~\ref{fig:mcfit-incl}, showing a broad peak towards higher
inclinations.  The median value and its asymmetric errors are
$i=46\degr^{+\:\,9\degr}_{-11\degr}$ (68\% C.L.).
%
%
%%%%%%%%%%%%%%%%%%%%%%%%%%%% FIGURE %%%%%%%%%%%%%%%%%%%%%%%%%%%%%%%%%%
%

%%%%%%%%%%%%%%%%%%%%%%%%%%%%%%%%%%%%%%%%%%%%%%%%%%%%%%%%%%%%%%%%%%%%%
\subsection{Non-detection of Shapiro delay \label{shapiro}}

The propagation of radio signals from pulsars in binary systems is
affected by a general-relativistic time delay of the signals in the
gravitational field of the companion star. For a pulsar in a circular
orbit this ``Shapiro delay'' (Shapiro 1964\nocite{sha64}) is given by
\begin{equation}
\Delta t = -2\, m_2\; T_{\odot}\, \ln[1-\sin i\, \sin(\Phi-\Phi_0)]\  ,
\end{equation}
where $\Phi$ is the orbital phase in radians, $\Phi_0$ is the phase of
the ascending node, and $T_{\odot}=(G\,M_{\odot}/c^3)$.  In practice,
Shapiro delay is conveniently expressed in terms of two observables,
the ``range'' $r = m_2\,T_{\odot}$ and the ``shape'' $s = \sin i$
(Ryba \& Taylor 1991\nocite{rt91a}), the post-Keplerian orbital
parameters which allow a determination of the companion mass, $m_2$,
and the orbital inclination, $i$. However, for small inclination
angles the variation of $\Delta t$ over the orbit is nearly sinusoidal
and cannot be separated from a small variation in $x$. For edge-on
orbits ($i\approx 90\degr$), $\Delta t$ peaks at $\Phi-\Phi_0=\pi/2$,
where the pulsar is behind the companion. Here, the covariance with
the Keplerian parameters breaks, and a measurement of Shapiro delay
(and hence $r$ and $s$) is possible. Because high inclination orbits
are relatively rare, Shapiro delay has been detected in only five
pulsar-white dwarf binaries, PSR B1855+09 (Ryba \& Taylor 1991), PSR
J1713+0747 (Camilo et al.\ 1994a\nocite{cfw94}), PSR J0437$-$4715 (van
Straten et al.\ 2001\nocite{vbb+01}), \object{PSR J1909$-$3744}
(Jacoby et al.\ 2003\nocite{jbv+03}), \object{PSR J1640+2224}
(L\"ohmer et al.\ 2004\nocite{llw+04}), and perhaps in \object{PSR
J0751+1807} (Nice et al.\ 2003\nocite{nss03}).

\begin{figure*}[ht]
\centering
\psfig{file=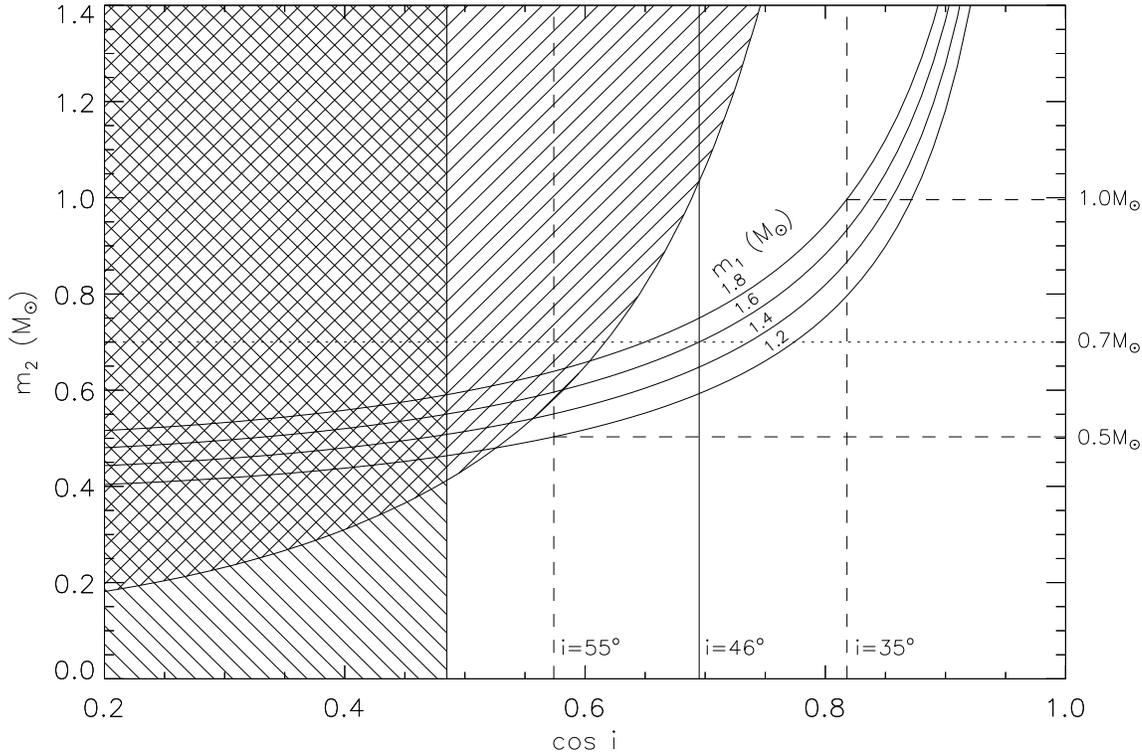,width=15cm}
\caption{\label{fig:mass-incl} Constraints on orbital inclination,
$i$, and WD mass, $m_2$, from $\dot x$-measurement and limits on
Shapiro delay. Points inside the hatched rectangle are excluded by
measured $\dot x$, points inside the curved hatched area are excluded
by the non-detection of Shapiro delay. From the mass function we
calculated the contours for constant neutron star masses, indicated by
the solid curves.  The vertical lines denote the median value
($i=46\degr$, solid) and its 1$\sigma$ errors (at $35\degr$ and
$55\degr$, dashed) for the distribution of inclination angles derived
from the observed $\dot x$. From these uncertainties we derive the
companion mass limits $0.5\, M_\odot\leq m_2\leq 1.0\, M_\odot$
displayed by the horizontal dashed lines. Photometric studies of the
WD companion (see Sect.~\ref{wdcool}) result in a lower mass limit of
$0.7\,M_\odot$ (horizontal dotted line). }
\end{figure*}

We did not detect Shapiro delay in the timing data of PSR
J2145$-$0750.  However, we can use the non-detection to place limits
on allowed values for $i$ and $m_2$.  Following a procedure described
by Nice et al.\ (2001), we tested a 2D grid of timing models, that
include the Shapiro parameters in the range of $0\le\sin i\le 1$ and
$0\le m_2\le 1.4$\,\msun, allowing all other parameters to vary.
The resulting $\chi^2$ values are mostly very similar to a model
without Shapiro delay. However, for models with high inclination
angles and/or high companion masses we obtain significantly higher
values for $\chi^2$. The failure of these models means that one would
have detected Shapiro delay had this been the true values of $i$
and $m_2$. In Fig.~\ref{fig:mass-incl}, the parameter space excluded
by models with $\Delta\chi^2>9$ (i.e., 3$\sigma$) is shown.

%%%%%%%%%%%%%%%%%%%%%%%%%%%%%%%%%%%%%%%%%%%%%%%%%%%%%%%%%%%%%%%%%%%%%
\subsection{Polarimetry\label{pol}}

During the recycling process of MSPs the neutron star is spun-up by
mass transfer from the companion star which is expected to cause an
alignment of pulsar spin and orbital angular momentum of the binary
system (Phinney \& Kulkarni 1994\nocite{pk94}).  Assuming that
this holds for PSR J2145$-$0750 it is then possible to constrain the
orbital inclination from the known geometry of the pulsar
magnetosphere.  We therefore observed PSR J2145$-$0750 at 1410~MHz
using the EBPP in polarization mode and derived a position angle that
is similar to the findings of Sallmen (1998\nocite{sal98}) and Stairs
et al.  (1999\nocite{stc99}). The profile of PSR J2145$-$0750 shows at
least three components, with a bridge of emission between the two
strongest components. Unfortunately, the linear intensity at 1410~MHz
is very weak (but see also Xilouris et al.\ 1998\nocite{xkj+98};
Sallmen 1998), and the accompanying position-angle swing is extremely
complex, with multiple orthogonal mode changes (Stairs et
al. 1999). Applying a rotating vector model (RVM) (Radhakrishnan \&
Cooke 1969\nocite{rc69a}) to the observed position angle usually leads
to high uncertainties in the RVM fit.  We fitted a RVM curve to the
position angle of the whole profile (including the precursor) and
found a best fit for a magnetic inclination of $\alpha\sim 52\degr$
and an impact angle of $\sigma\sim -3\degr$. These values are
considerably different from those found by Sallmen (1998) which is
probably due to the inclusion of the precursor in our fit leading to a
longer baseline in pulse longitude.  Even though the position angle is
poorly constrained leading to considerable errors in the RVM fits,
applying our best-fit values the angle between the pulsar rotation
axis and the line-of-sight results in $\alpha+\sigma\sim 49\degr$. If
one assumes that the pulsar rotation axis is aligned with the orbital
momentum, one derives $i=\alpha+\sigma$ (or
$i=180\degr-(\alpha+\sigma)$), i.e.  $i\sim 49\degr$.

One generally needs to be careful in interpreting the position angles
observed in recycled pulsars as being the reflection of the underlying
geometry. The position angle swings are typically much flatter and
often do not resemble the expected S-like swing (e.g.\ Xilouris et
al.\ 1998; Stairs et al.\ 1999).  Given this caveat in mind, it is
interesting to note that the fitted inclination angle agrees very well
with the median value of $i=46\degr$ obtained from our $\dot x$
measurement.  We conclude that the analysis of the pulsar
magnetosphere by polarimetry results in a consistent picture of the
orbital inclination, strongly supporting our earlier findings.

%
%%%%%%%%%%%%%%%%%%%%%%%%%%%% FIGURE %%%%%%%%%%%%%%%%%%%%%%%%%%%%%%%%%%
%

%%%%%%%%%%%%%%%%%%%%%%%%%%%%%%%%%%%%%%%%%%%%%%%%%%%%%%%%%%%%%%%%%%%%%
\subsection{Companion mass\label{mass}}

As demonstrated, the $\dot x$ measurement and the limits on Shapiro
delay for PSR J2145$-$0750 can be used to put constraints on the
orbital inclination, $i$, and the companion mass, $m_2$, as shown in
Fig.~\ref{fig:mass-incl}. From the observed $\dot x$ we derive an
upper limit on the inclination angle of $i<61\degr$, so that all
points with $\cos i<0.48$ are excluded as indicated by the hatched
rectangle in the figure. The curved hatched area is the parameter
space excluded due to the non-detection of Shapiro delay in the
binary. From the mass function of PSR J2145$-$0750 we derive the
displayed contours for constant neutron star masses.

We cannot put any constraints on the companion mass from evolutionary
considerations of the binary as the PSR J2145$-$0750 system most
probably went through a phase of highly unstable mass accretion (van
den Heuvel 1994). Thus, an unique relation between the orbital period,
$P_{\rm b}$, and the secondary mass, $m_2$, as found for LMBPs, is not
applicable for this system. However, from the distribution of allowed
inclination angles (see Sect.~\ref{xdot}), and the contours for
constant pulsar masses we can estimate a range of allowed companion
masses. Using recent results of neutron star mass measurements
(Thorsett \& Chakrabarty 1999\nocite{tc99}; Stairs 2004\nocite{sta04}
and references therein) we apply a range of $1.2\, M_\odot\leq m_1\leq
1.8\, M_\odot$ for the pulsar mass (for a discussion see
Sect.~\ref{concl}).  As shown in Fig.~\ref{fig:mass-incl}, for
inclination angles within their 1$\sigma$ uncertainties we derive a
range of allowed companion masses of $0.5\, M_\odot\leq m_2\leq 1.0\,
M_\odot$.

%%%%%%%%%%%%%%%%%%%%%%%%%%%%%%%%%%%%%%%%%%%%%%%%%%%%%%%%%%%%%%%%%%%%%
\section{The nature and cooling age of the white dwarf companion
\label{wdcool}}

From the timing analysis we derive a range of companion masses of
$0.5\, M_\odot\leq m_2\leq 1.0\, M_\odot$, and a global minimum of
$m_{\rm 2, min}=0.48\, M_\odot$ for an extremely low pulsar mass of
$m_1=1.2\, M_\odot$ (see Sect.~\ref{concl}). The upper mass limit for
a helium (He)-core WD with solar metallicity is $M_{\rm He-WD,
max}\approx 0.45\, M_\odot$. Hence we conclude that the companion is a
WD with carbon/oxygen (CO) core in the intermediate WD mass
range. This result is consistent with a CE evolution of the binary
system as proposed by van den Heuvel (1994). Using the derived mass
range of the WD companion and its photometrically determined effective
temperature it is possible to determine an age of the binary
system. This age is based on WD cooling models coming from stellar
evolution theory and can be compared with the characteristic pulsar
age.

Optical observations of the WD companion with the HST from Lundgren et
al.\ (1996) yielded $m_{\rm V}=23.7\pm0.1$ and $m_{\rm
I}=23.0\pm0.1$. From these measurements the authors derived an
effective temperature of $T_{\rm eff}=5800\pm300\, {\rm K}$ using the
color-temperature calibration from Bergeron et al.\
(1995\nocite{bsw95}) and the bolometric correction from Monet et al.\
(1992\nocite{mdv+92}).  With the same observational data from Lundgren
et al.\ (1996), Hansen \& Phinney (1998)\nocite{hp98b} estimated
$T_{\rm eff}=6235\pm970\, {\rm K}$ based on the $V-I$ broad-band
colors of Bergeron et al.\ (1995). As conservative limits, we thus use
a range of effective temperatures for the companion between 5250 and
7200 K.

\begin{figure}
\centering 
\psfig{file=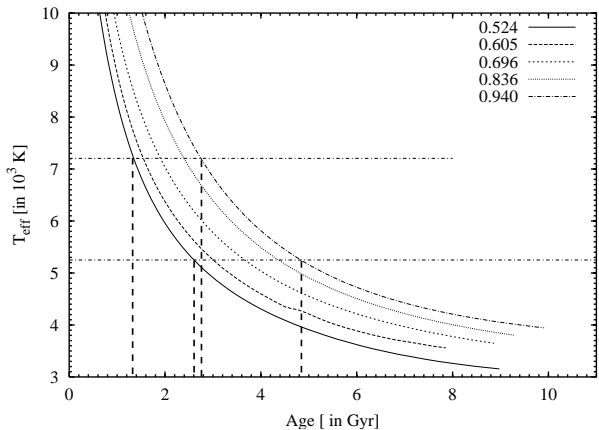,angle=-90,width=8.3cm}
\caption{ \label{wdtteff} Age -- temperature diagram with cooling
curves from evolutionary models of Bl\"ocker (1995) and Bl\"ocker
et al.\ (1997) for CO-core WDs ($0.52\,M_\odot < M_{\rm WD} <
0.94\,M_\odot$). The labels give the WD masses in $M_\odot$, and the
straight horizontal lines indicate the possible range of effective
temperatures for the WD companion of PSR J2145$-$0750 according to the
studies of Lundgren et al.\ (1996) and Hansen \& Phinney (1998),
i.e.\ 5250~K $\leq T_{\rm eff} \leq$ 7200~K. The thick dashed vertical
lines mark the intersections of these lines with the cooling curves
from the least and the most massive CO-core model ($M_{\rm
WD}=0.524$ and $0.940\, M_\odot$, respectively) to indicate the
corresponding age range.  
}
\end{figure}

Fig.\ \ref{wdtteff} shows an age -- $T_{\rm eff}$ diagram with
evolutionary tracks of CO-core WD models from Bl\"ocker
(1995\nocite{blo95b}) and Bl\"ocker et al.\ (1997\nocite{bhd+97}). The
horizontal lines indicate the observational range for $T_{\rm
eff}$. The curve at the very left of the diagram belongs to the
CO-core WD model with the lowest mass, followed by curves of
increasing core mass while going to right. The WD models with masses
of 0.524 and 0.940 $M_\odot$ roughly bracket the mass range for the
companion derived from the timing analysis.  As the vertical
dashed lines in Fig.~\ref{wdtteff} illustrate, a WD with $m_{\rm
WD}=0.524\, M_\odot$ would have an age of $1.3\dots2.6$ Gyr, whereas a
massive WD with $m_{\rm WD}=0.940\, M_\odot$ would have an age of
$2.8\dots4.9$ Gyr.

%
%%%%%%%%%%%%%%%%%%%%%%%%%%%% FIGURE %%%%%%%%%%%%%%%%%%%%%%%%%%%%%%%%%%
%%%%%%%%%%%%%%%%%%%%%%%%%%%%%%%%%%%%%%%%%%%%%%%%%%%%%%%%%%%%%%%%%%%%%%

Using the photometric measurements of Lundgren et al.\
(1996\nocite{lfc96}) in the $B$, $V$, and $I$ bands together with our
parallax distance for \object{PSR J2145$-$0750}, the mass, age, and
effective temperature of the WD companion can be further
constrained. With $d=500^{+210}_{-120}$~pc and the apparent magnitudes
$m_{\rm B}=23.9$, $m_{\rm V}=23.7$, and $m_{\rm I}=23.0$ from Lundgren
et al.\ (1996\nocite{lfc96}) we find absolute magnitudes $M_{\rm
B}=15.4^{+0.6}_{-0.8}$, $M_{\rm V}=15.2^{+0.6}_{-0.8}$, and $M_{\rm
I}=14.5^{+0.6}_{-0.8}$.  Applying the bolometric corrections $BC_{\rm
I}= 0.5$ from Monet et al.\ (1992) and $BC_{\rm V}= -0.2$ from
Bergeron et al.\ (1995\nocite{bsw95}), from both $M_{\rm V}$ and
$M_{\rm I}$ we obtain a bolometric brightness of $M_{\rm
bol}=15.0^{+0.6}_{-0.8}$, which finally gives a WD luminosity of $\log
(L/L_\odot) = -4.1^{+0.3}_{-0.2}$.  This value can be compared with
the luminosity predictions from the WD cooling models from Bl\"ocker
(1995\nocite{blo95b}) and Bl\"ocker et al.\ (1997\nocite{bhd+97}) for
$T_{\rm eff}=5250 \dots 7200$ K. The model parameters are summarized
in Table \ref{tabwd2}.
%}
%%%%%%%%%%%%%%%%%%%%%%%%%%%%%%%
%%%% Tab.: WD model parameters
%%%%%%%%%%%%%%%%%%%%%%%%%%%%%%%
\begin{table}
\begin{center}
\caption{Cooling properties of the CO-core WD companion of \object{PSR J2145$-$0750}.
All values are taken from WD evolutionary models of 
\cite{blo95b} and \cite{bhd+97}.
\label{tabwd2}
}
\vspace{0.1cm}
\begin{tabular}{llll}
\hline\hline \multicolumn{4}{c}{\it Model parameters of the WD
companion}\\\hline 
$M_{\rm WD}/M_\odot$& $\log (L/L_\odot)$ & $T_{\rm
eff}/$K & $\tau_{\rm cool}/$Gyr\\\hline
0.524               & -3.348 & 7200 & 1.35\\
0.524               & -3.913 & 5250 & 2.60\\\hline
0.605               & -3.448 & 7200 & 1.50\\
0.605               & -3.995 & 5250 & 3.00\\\hline
0.696               & -3.532 & 7200 & 1.90\\
0.696               & -4.087 & 5250 & 3.70\\\hline
0.836               & -3.661 & 7200 & 2.40\\
0.836               & -4.285 & 5250 & 3.40\\\hline
0.940               & -3.825 & 7200 & 2.80\\
0.940               & -4.850 & 5250 & 4.85\\\hline
\end{tabular}
\end{center}
\vspace*{-2mm}

\end{table}
%%%%%%%%%%%%%%%%%%%%%%%%%%%%%%%%

As can be seen from Tab.\ \ref{tabwd2}, using $\log (L/L_\odot) =
-4.1$ as an additional constraint a CO-core WD with $m_2\la
0.7\,M_\odot$ can be ruled out as companion for \object{PSR
J2145$-$0750}, since WDs with lower masses would be too
luminous. Here, a WD with $m_2\approx 0.7\,M_\odot$ would just have
$\log (L/L_\odot) = -4.1$ in accordance with the observations, and the
cooling age would be $\tau_{\rm cool}=3.7$ Gyr.  On the other hand, a
WD with $m_2=0.94\,M_\odot$ would be too faint at a temperature of
$T_{\rm eff}=5250$ K. Using $\log (L/L_\odot) = -4.1$ as constraint,
we obtain $\tau_{\rm cool} =3.4$ Gyr from the $0.94\,M_\odot$ cooling
model, which in turn leads to a new upper limit for the effective
temperature of $T_{\rm eff}=6400$~K (see Fig.\ \ref{wdtteff}).
Therefore, from the overall comparison of observational constraints
and WD models, we conclude that the PSR J2145$-$0750 companion has a
mass of $0.7\, M_\odot\leq m_2\leq 1.0\, M_\odot$, an effective
temperature of $T_{\rm eff}=5750\pm600$ K, and a cooling age of
$\tau_{\rm cool}=3.6(2)$ Gyr.

The WD mass range derived here is in accordance with the results from
our timing analysis (see Sect.\ \ref{mass}), but provides a tighter
lower limit of $0.7\,M_\odot$. On the other hand, the results on the
WD properties are also in agreement with previous studies from
Lundgren et al.\ (1996) and Hansen \& Phinney (1998). Taking the mass
range $0.7\, M_\odot\leq m_2\leq 0.9\, M_\odot$, we can also compare
our results with those obtained from the recent WD models of Richer et
al.\ (2000\nocite{rhl+00}).  Using their models, we obtain $T_{\rm
eff}=5700\pm350$ K and $\tau_{\rm cool}=4.4(1)$ Gyr in line with our
results.

%%%%%%%%%%%%%%%%%%%%%%%%%%%%%%%%%%%%%%%%%%%%%%%%%%%%%%%%%%%%%%%%%%%%%
\section{Discussion \& Conclusions \label{concl}}

Our timing observations of the PSR J2145$-$0750 binary system
constrain the orbital inclination angle to be $i<61\degr$, with a
median value of $i=46\degr$. Polarimetric studies of the pulsar
magnetosphere lead to a consistent result but cannot provide stronger
constraints on the orbital inclination. From a statistical analysis of
allowed inclination angles we are able to derive proper mass limits
for the pulsar companion if we apply a realistic estimate for the
pulsar mass.  Neutron stars are believed to have mass close to the
Chandrasekhar value of $\sim$1.35\,\msun. Thorsett \& Chakrabarty
(1999) reviewed all neutron star mass measurements and found a
remarkably narrow distribution of $1.35(4)\,M_\odot$. Recent
observations revealed neutron stars that have lower masses, e.g.\
\object{PSR J0737$-$3039B} with $1.250(5)\,M_\odot$ in the newly
detected double-pulsar system (Lyne et al.\ 2004\nocite{lbk+04}), and
\object{PSR J1141$-$6545} with $1.30(2)\,M_\odot$ in a NS-WD system
(Bailes et al.\ 2003\nocite{bok+03}). New timing observations showed
that recycled pulsars in NS-WD binaries tend to have masses that are
significantly higher than the canonical value of $1.35\,M_\odot$ (Nice
2003\nocite{nic03}), which is generally explained by a long phase of
extended mass transfer from the companion to the NS.  This also holds
true for \object{PSR J0621+1002} with a pulsar mass of
$\sim$1.7\,\msun\ (Splaver et al.\ 2002\nocite{sna+02}), an IMBP
system with properties similar to PSR J2145$-$0750.  The currently
known NS masses in recycled pulsar-WD systems lie in the range $1.5 -
1.7\,M_\odot$, with a median value of $\sim$1.6\,\msun\ (ignoring PSR
J0751+1807 due to the large uncertainties in its mass measurements;
Stairs 2004 and references therein). Taking into account the
considerable uncertainties, we conservatively choose a range of $1.2\,
M_\odot\leq m_1\leq 1.8\, M_\odot$ for the pulsar mass of PSR
J2145$-$0750. As discussed in Sect.~\ref{mass}, this results in
allowed companion masses of $0.5\, M_\odot\leq m_2\leq 1.0\,
M_\odot$. We hence conclude that the pulsar companion is a CO-core WD
in the intermediate WD mass range. For the median pulsar mass of
$\sim$1.6\,\msun\ and the observed median inclination angle of
$46\degr$ we obtain a median companion mass of $\sim$0.7\,\msun.  

Using the measured parallax distance and the constraints coming from
photometrical observations of the WD companion of PSR J2145$-$0750,
the WD properties can be further constrained. Our analysis reveals
that the companion has a mass in the range $0.7\, M_\odot\leq m_2\leq
1.0\, M_\odot$, a cooling age of $\tau_{\rm cool}=3.6(2)$~Gyr, and an
effective temperature of $T_{\rm eff}=5750\pm600$~K, in good agreement
with results from previous studies (Lundgren et al.\ 1996, Hansen \&
Phinney 1998, Richer et al.\ 2000). In particular, the WD model for a
companion mass of $0.7\,M_\odot$, the median value derived from the
timing analysis, exactly reproduces the measured WD luminosity for
$T_{\rm eff}=5250$~K.

The WD cooling age is roughly a factor of 3 lower than the pulsar's
characteristic age of 10.4 Gyr, which is generally considered as a
rough upper limit of the system age (Camilo et al.\
1994b\nocite{ctk94}). Under the assumption that the pulsar's magnetic
field does not decay, we can use the WD cooling age to estimate the
initial spin period $P_0$ of PSR J2145$-$0750, by inverting
Eq.~\ref{pulsarage}. The Doppler contributions are subtracted as we
apply the intrinsic spin-down rate $\dot P^{\rm intr}$.  For the
derived system age of $\tau=\tau_{\rm WD}=3.6(2)$ Gyr and typical
braking indices $n=2...3$ (Lyne 1996\nocite{lyn96} and references
herein) we hence obtain an initial spin period of $P_0=13.0(5)$~ms,
which is very close to the current period. Theories regarding the
spin-up of recycled pulsars to millisecond periods (Smarr \& Blandford
1976\nocite{sb76}; Ghosh 1995\nocite{gho95b} and references therein)
predict that the initial period should be equal to the equilibrium
spin period of the neutron star in the X-ray binary phase, accreting
at the Eddington rate $\dot M_{\rm Edd}$. For PSR J2145$-$0750 we find
$\dot M/\dot M_{\rm Edd}=0.004(1)$. Hence the final accretion rate of
the pulsar was highly sub-Eddington, which was also observed in other
IMBP systems (Lundgren et al.\ 1996). This is not surprising as the
unstable CE phase of the binary was too short to spin up the pulsar to
the equilibrium spin period. Instead, as proposed by van den Heuvel
(1994), the initial period may have resulted from sub-Eddington
accretion from the stellar wind of the giant star before reaching the
asymptotic giant branch phase.

The prospects for measuring the pulsar mass of the PSR J2145$-$0750
system are quite low. As the pulsar companion is bright enough in the
optical, the system masses could in principle be determined by a
spectroscopic study (see van Kerkwijk et
al.~1996\nocite{vbk96}). However, given the composition of the
companion, suitable lines might not be present.  Similarly,
post-Keplerian effects beyond Shapiro delay will not be detectable in
the foreseeable future.  As mentioned in Sect.~\ref{secchange},
relativistic effects such as orbital period decay or periastron
advance are fully absorbed by the redefinition of the binary period
and therefore not observable in this low-eccentricity binary. The
precision of the Shapiro delay measurement (or limit) cannot be
improved drastically with existing radio telescopes. It will need
telescopes like the Square-Kilometre-Array (SKA) to provide a large
sample of precision mass measurements.

%%%%%%%%%%%%%%%%%%%%%%%%%%%%%%%%%%%%%%%%%%%%%%%%%%%%%%%%%%%%%%%%%%%%%
\begin{acknowledgements}

We are very grateful to all staff at the Effelsberg and Jodrell Bank
observatories for their help with the observations. We thank Oleg
Doroshenko, Christoph Lange, and Norbert Wex for their assistance in
the Effelsberg timing project. We would also like to thank the referee
for useful suggestions that helped to improve the text. OL was was
partially funded during this work by the European Commission, Marie
Curie Training Site programme, under contract no. HPMT-CT-2000-00069.

\end{acknowledgements}

%%%%%%%%%%%%%%%%%%%%%%%%%%%%%%%%%%%%%%%%%%%%%%%%%%%%%%%%%%%%%%%%%%%%%
\bibliographystyle{aa}
%\bibliography{journals,modrefs,psrrefs,loehmer,crossrefs}

\end{document}